\documentclass[11pt,twocolumn]{scrartcl}

\usepackage{casa_conf}	
\usepackage{graphicx}	

\usepackage[inkscapearea=page]{svg}
\usepackage{amsmath}
\usepackage{amssymb}  
\usepackage{subfigure}
\usepackage{array}
\usepackage{booktabs, boldline, makecell}
\usepackage{color,soul}
\usepackage{titlesec}
\titlespacing*{\section}{0pt}{1.0\baselineskip}{\baselineskip}
\titlespacing*{\subsection}{0pt}{1.0\baselineskip}{\baselineskip}

\title{Synthesizing Human Faces using Latent Space Factorization and Local Weights (Extened Version)}

\author{Minyoung Kim\\
        Ewha Womans University\\
        minyoung.mia.k@ewhain.net\\
        https://minyoung-mia-kim.github.io
        \and
        Young J. Kim\\
        Ewha Womans University\\
        kimy@ewha.ac.kr\\
        http://home.ewha.ac.kr/~kimy/}

\begin{document}

\maketitle

\begin{figure*} [ht]
\centering
\includegraphics [width=7.5cm]{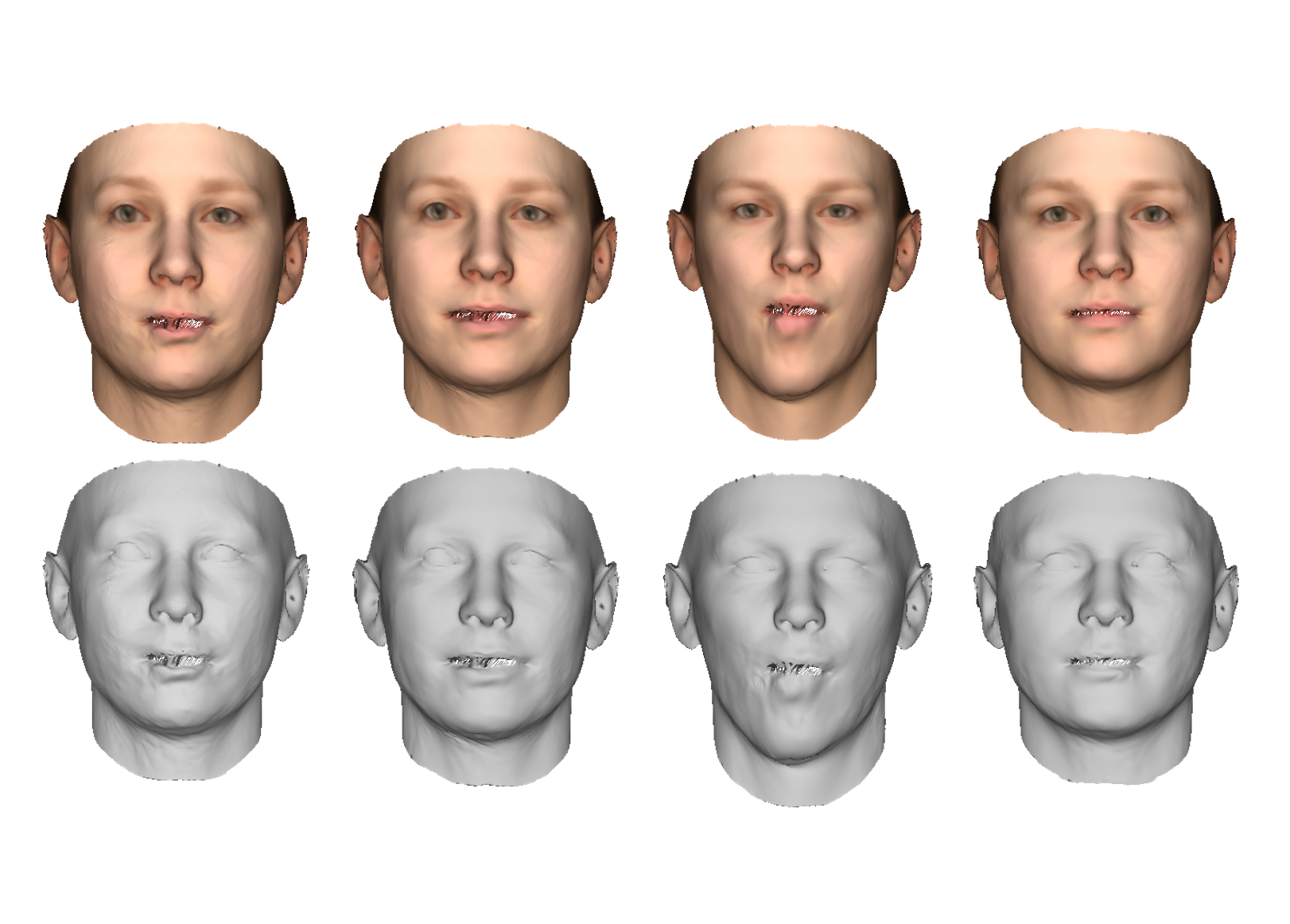}
\includegraphics [width=7.5cm]{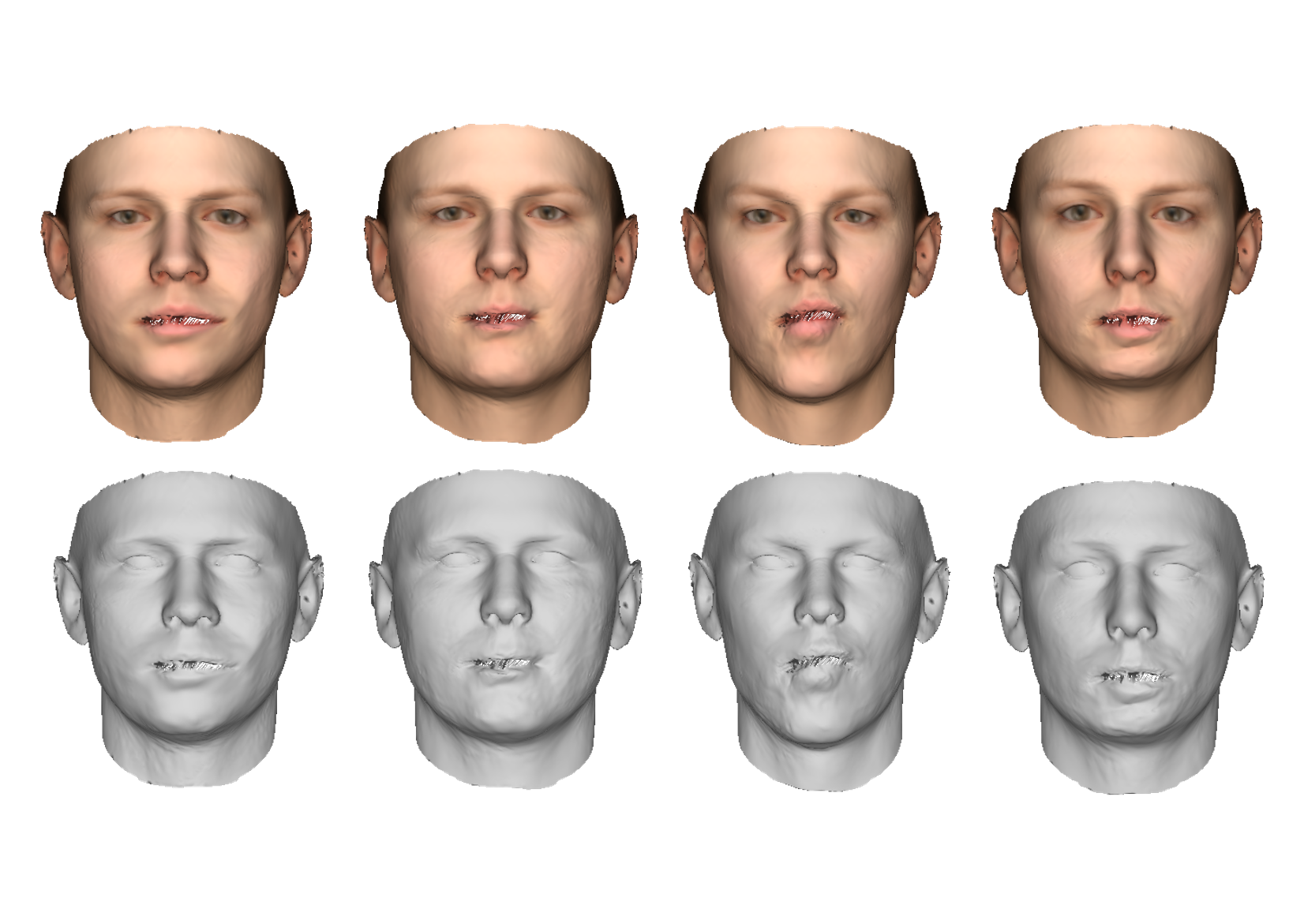}
\vspace{-20pt}
\caption{
Facial synthesis samples created by our generative model}
\label{fig:syn}
\vspace{-10pt}
\end{figure*}

\begin{abstract}
We propose a 3D face generative model with local weights to increase the model’s variations and expressiveness. The proposed model allows partial manipulation of the face while still learning the whole face mesh. For this purpose, we address an effective way to extract local facial features from the entire data and explore a way to manipulate them during a holistic generation. First, we factorize the latent space of the whole face to the subspace indicating different parts of the face. In addition, local weights generated by non-negative matrix factorization are applied to the factorized latent space so that the decomposed part space is semantically meaningful. We experiment with our model and observe that effective facial part manipulation is possible, and that the model’s expressiveness is improved.
\end{abstract}
\linebreak
\linebreak

\section{Introduction}
Various methods have been studied to develop three-dimensional(3D) geometric models to generate human faces. 
Its importance has increased lately due to the progress in virtual reality, especially virtual humans~\cite{hu2017avatar}. However, modeling a human face still needs a tremendous human effort. Many researchers have proposed new approaches to address this difficulty. Among them, the learning-based method exhibits notable advancements recently. 

With the advent of the generative adversarial networks (GAN), attention to the generation model is increasing, and related research using deep learning is being actively pursued. However, most existing works are focused on a holistic generative approach to generate all parts at once and lack part details and manipulation.


Previous part-based generative models exploit explicit segmentation data or labels for training their model to learn the structure of the object parts or use several part decoders\cite{wang2018global,schor2019componet,dubrovina2019composite}. However, existing 3D facial mesh datasets barely have pre-segmented data.
Thus, we investigate an effective way to extract or present localized features from the whole data. Toward this goal, mesh segmentation might be one of the possible solutions. However, since human faces are often smooth, it is a challenge to segment the facial mesh explicitly. To bypass this, we exploit a generative approach that does not require additional segmentation data and makes the whole learning model simple. Furthermore, we explore a way for part control while exploiting holistic generation by learning localized features. 

In this paper, we propose a locally weighted 3D face generative model to increase variations and expressiveness of the model. Our approach can generate a rich variety of 3D face models beyond the training data using part manipulation with latent factorization. With a part-based representation of the data, our model is simpler and more straightforward than others and does not require any semantic segmentation labels. 

Our main contributions are: (1) Locally weighted generative autoencoder for generating a whole human face geometric model; (2) End-to-end learning to learn local features without explicit facial feature segmentation data; (3) Experimentation and demonstration of the proposed model’s performance in terms of generation and part manipulation.


As a basis model, we leverage Ranjan et al.~\cite{ranjan2018generating}’s autoencoder with latent space factorization and apply local weights that partially influence the model during training. Latent factorization enables manipulation of the local part of the face, and local weights make decomposed part spaces more semantically meaningful without additional segmentation labels. 
We also evaluate the performance of the proposed model in terms of part modification, part combination, and ablation tests to show the effect of each model component on the results. The majority of the materials contained in this paper is based on the same author's dissertation \cite{Thesis20}.

\section{Related Work}
\subsection{3D Face Representations and Local models}
Blanz and Vetter~\cite{blanz1999morphable} introduced the first 3D face morphable models (3DMMs), which are statistical models of global 3D face shapes and textures. They employed principal component analysis (PCA) to construct principal components to express facial shape and texture. More recently, Booth et al.~\cite{booth20163d} proposed the first largest scale morphable model, the large scale face model (LSFM)~\cite{booth20163d}, constructed from 9663 distinct facial identities. Paysan et al.’s Basel face model (BFM)~\cite{paysan20093d} also has been widely used. However, the 3DMMs are limited to representation of high-frequency details and form a latent model space due to their linear bases and training data. 

There exist attempts to generate a new face with face segmentation and a local model to increase the model’s expressiveness and achieve fine-scale modeling. Blanz and Vetter~\cite{blanz1999morphable} demonstrated region-based modeling with 3DMMs by manually dividing the face into regions that can be learned by the PCA models. Tena et al.~\cite{tena2011interactive} presented region-based linear face modeling with automatic segmentation by clustering. Tran et al.~\cite{tran2019towards} also proposed non-linear 3DMMs with a global and local-based network to extract features of the global face structure and face part details simultaneously. Recently, Ghafourzadeh et al.~\cite{ghafourzadeh2020part} proposed a part-based approach that conducts part-based facial models using PCA. This model results in a locally edited face by applying an anthropometric measurement. 



\subsection{Part-Based Shape Generative Models}
With the advantages of CNNs for hierarchical feature extraction, many generative models also capitalize on its benefits to progress shape modeling. 

Wang et al.~\cite{wang2018global} proposed holistic voxel-based generative adversarial networks called global-to-local GAN and part refiner. 
They showed better shape variety and distribution than a plain three-dimensional GAN. 
CompoNet~\cite{schor2019componet} presented a part-based generative neural network for shapes. 
They proved that the part-based model encourages the generator to create new data unseen in the training set. Dubrovina et al.~\cite{dubrovina2019composite} handled the composition and decomposition of each part as a simple linear operation on the factorized embedding space.
They used projection matrices to split full object encodings into part encodings. 
The proposed decomposer-composer network can perform meaningful part manipulations and high-fidelity 3D shape generation. 

To composite each part, both~\cite{schor2019componet} and~\cite{dubrovina2019composite} compute per-part affine transformation. 
Our model does not utilize spatial transformer networks~\cite{jaderberg2015spatial} nor computes affine transformation to combinate each part of the data. 
We pursue a holistic generation approach but also allow part manipulation without explicit segmentation. Therefore, we do not need to worry about the artifacts when the model combines each part into a whole shape.

\subsection{Feature Matrix Factorization}
Some feature factorization methods interpret data more semantically since they can decompose the data into a part-based representation. Among those methods, non-negative matrix factorization (NMF) is a robust feature factorization method to represent data as part-based ones. Lee and Seung ~\cite{lee1999learning} popularized NMF by showing its interpretability for the part-based representation of facial images. Koppen et al.~\cite{koppen2016extending} extended NMF to 3D registered images. McGraw et al.~\cite{mcgraw2016sparse} presented 3D segmentation based on NMF and produce meaningful results. For its application, Li et al.~\cite{li2020sketch} proposed the concept of sketch as an input of GANs, which is the noise transformed to the basis matrix in NMF that has the underlying features of the raw data.

 By utilizing a part-based representation, Collins et al.~\cite{collins2020editing} performed local and semantically-aware changes through a global operation on the 2D image domain. They applied spherical k-means clustering \cite{buchta2012spherical} on the last feature map to identify features that are semantically meaningful. Collins and Süsstrunk~\cite{collins2019deep} demonstrated localized features using NMF. They applied NMF to the last feature map, where the semantic features are encoded. By factorizing the feature map, they can decompose an input image into several semantic regions.

Inspired by these works, we apply NMF to 3D faces mesh to obtain local features of data and supply them to the holistic generative model.

\section{Mesh Convolution Neural Networks}

We choose to represent 3D faces with triangular mesh due to its efficiency. 
Among previous approaches for applying mesh convolution operation, Ranjan et al.~\cite{ranjan2018generating} proposed CoMA employing fast Chebyshev filters ~\cite{defferrard2016convolutional} with a novel mesh pooling method. 
Since our model is largely based on this model, we briefly describe the convolution operation applied to mesh data.

\begin{figure*}[ht]
\centering
\includegraphics[width=\textwidth]{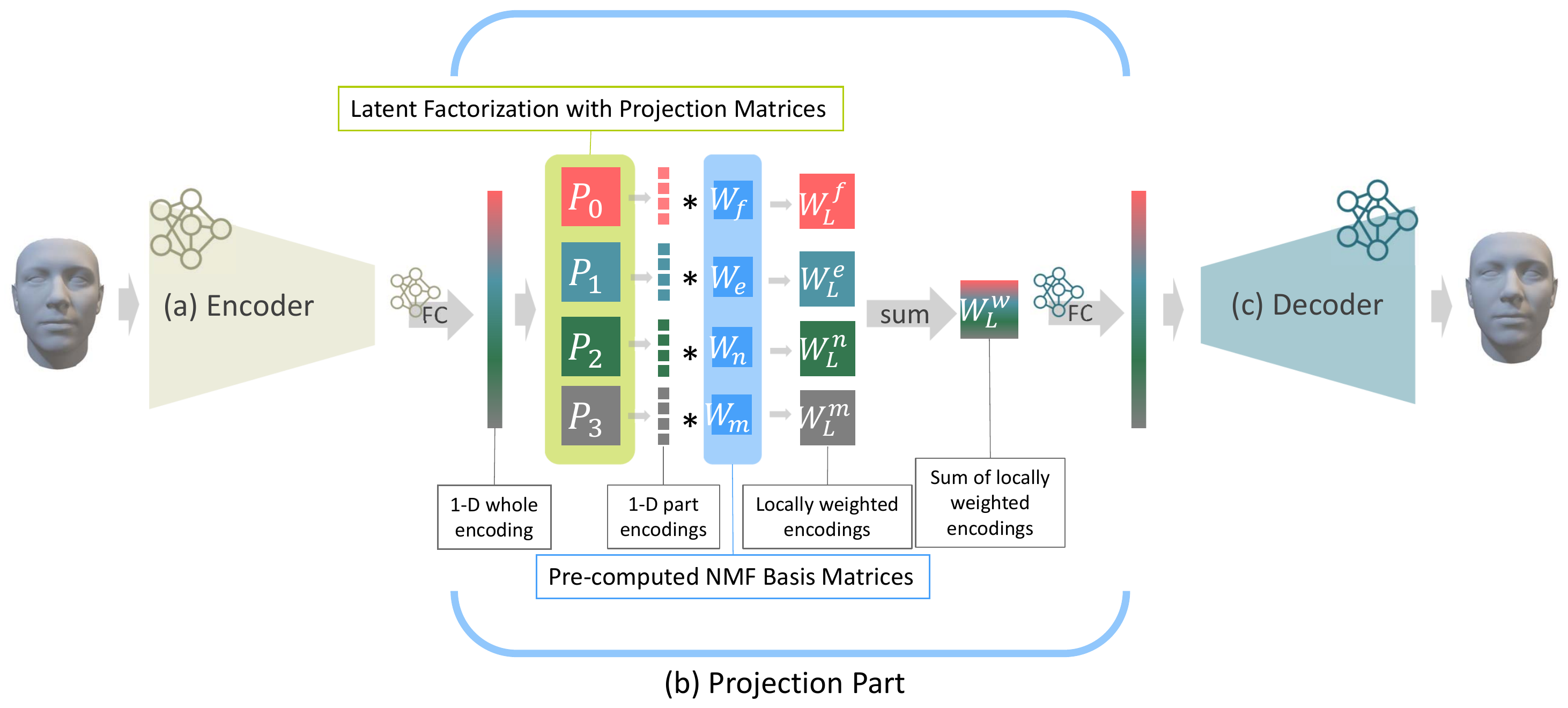}
\caption{
Locally weighted autoencoder architecture}
\label{fig:full_arch}
\end{figure*}

\subsection{Mesh Representation}
We represent a 3D face mesh as a set of vertices $V\in \mathbb{R}^{N\times 3}$ and edges $E$. The edges are represented by an adjacency matrix $A\in \{0,1\} ^{N\times N}$ where $A_{ij}=1 $ denotes where there is an edge connecting vertices $i$ and $j$, and $A_{ij}=0$ otherwise.

\subsection{Spectral Graph Convolution on Face Mesh}
Defferrard et al.~\cite{defferrard2016convolutional} use convolution on graphs with a frequency domain approach under the convolution theorem. The convolution in the spatial domain equals element-wise multiplication in the frequency domain. To convert the graph from the spatial domain to the frequency domain, Defferrard et al.~\cite{defferrard2016convolutional} first applied the graph Fourier transform ~\cite{chung1997spectral} to the input mesh. The graph Laplacian matrix is defined as $L= D-A$, where $D$ is a diagonal matrix, with $D_{i,i} = \mathbf{\Sigma}_jA_{ij}$. The Laplacian matrix is diagonalized by the Fourier basis $U\in \mathbb{R}^{N\times N}$ as $L={U\mathrm{\Lambda U}}^T$. Here, the columns of $U=[\ u_0,\ u_1,\ \ldots,\ u_{n-1}]$ are the orthogonal eigenvectors of $L$, and $\Lambda=diag\left(\left[\ \lambda_0,\ \lambda_1,\ \ldots,\ \lambda_{n-1}\right]\right)\in\mathbb{R}^{N\times N}$ is a diagonal matrix. Following the convolution theorem, the convolution operator $*$ can be defined in the Fourier space as the element-wise product $X\ast W_{spec}=U(U^T(X)\odot U^T(W_{spec}))$. Because of $U$, which is not sparse, this operation needs high computational costs. To address this problem, Defferrard et al.~\cite{defferrard2016convolutional} formulate spectral convolution with a filter $W_\theta$ using a recursive Chebyshev polynomial (\cite{defferrard2016convolutional}, \cite{hammond2011wavelets}). The filter $W_\theta$ is parametrized as a Chebyshev polynomial of order $k$ by
\begin{equation} \label{eq:W}
    W_\theta\left(L\right)\ =\sum_{k=0}^{K-1}{\theta_kT_k\left(\widetilde{L}\right)},
\end{equation}
where $\widetilde{L}=\frac{2L}{\lambda_{max}}-I_n$ is the scaled Laplacian matrix, and $\lambda_{max}$ is the maximum eigenvalue of the Laplacian matrix. The parameter $\theta\in\mathbb{R}^K$ is a vector of the Chebyshev coefficients, and $T_k\in\mathbb{R}^{N\times N}$ is the Chebyshev polynomial of order k, which is computed recursively as $T_k\left(x\right)=2xT_{k-1}\left(x\right)-\ T_{k-1}(x)$, with $T_0=1$ and $T_1=x$. For each convolution layer, the spectral graph convolutions are
\begin{equation} \label{eq:Y}
    \mathcal{Y}_j=\sum_{i=0}^{F_{in}}{W_{\theta_{i,j}}(L)}x_i,
\end{equation}
where $x_i$ is the $i$-th feature of the input $x\in\mathbb{R}^{N\times F_{in}}$, and $\mathcal{Y}_j$ is the $j$-th feature of the output $\mathcal{Y}\in\mathbb{R}^{N\times F_{out}}$. For each convolution layer, the spectral graph convolution has $F_{in}\times F_{out}$ vectors of the Chebyshev coefficient $\theta_{i,\ j}\in\mathbb{R}^K$ as trainable parameters.

\section{Locally Weighted Autoencoder}
Our model is based on the autoencoder\cite{ranjan2018generating} consisting of an encoder, projection, and a decoder, as illustrated in Fig.{~\ref{fig:full_arch}}. Although their work has shown a decent performance of reconstructing 3D faces, we take one step further to improve generation ability and controllability by using per-part manipulation. Utilizing the basic generation ability of Ranjan et al ~\cite{ranjan2018generating}'s model, we added two new methods: latent factorization and local weights.
The encoder and decoder learn how to compress and decompress the data, respectively. In between them, the projection part factorizes the latent space into the subspace and applies local weights to make the subspace semantically meaningful.
More details on local weights and latent space manipulation will be explained in the following section.

\subsection{Pre-Computed Local Weights from NMF}

We use a part-based representation to extract the local part structure without segmented data or labels. The representation is used as weights, which have each vertex’s influence on each divided facial part. 
 
To make a part-based representation of the whole data, we employ the NMF. This method finds a low-rank approximation of a matrix $V$, where $V \approx WH$, when $V,W$, and $H$ do not have non-negative values. Given a feature matrix, $V, W$ is a basis matrix that contains basis elements of $V$, and $H$ is a latent representation matrix. We call the matrix $W$ local weights. To express local features more efficiently, we applied sparse NMF{~\cite{potluru2013block}} enforcing sparsity on the column of $H$. This could improve the local separation of features{~\cite{mcgraw2016sparse}}. We compute this with a sparsity constraint value of 7.5. The computed local weights serve as the influence of each vertex on a specific area. We expect that local weights would make the part encodings more semantically meaningful.

\begin{figure}[thb]
\centering
\includegraphics[width=\linewidth]{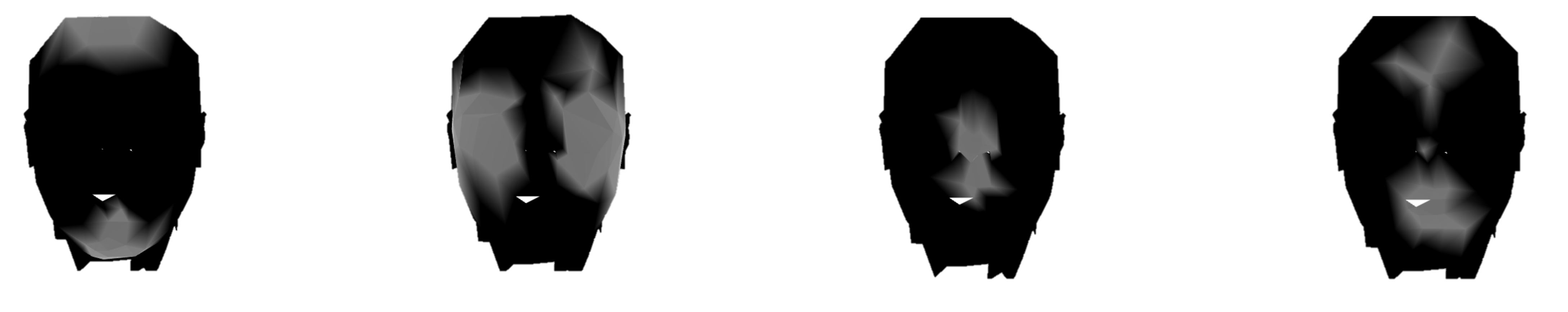}
\caption{Pre-computed sparse NMF's basis matrix}
\label{fig:lw}
\vspace{-10pt}
\end{figure}

Fig.~\ref{fig:lw} shows the visualization of the local weights. The bright area shows how much each vertex influences the facial area.

Before training the model, we compute a basis matrix $W$ and input a simplified template mesh vertices matrix $V\in \mathbb{R}^{P\times3}$  , where $P$ is the number of vertices with positions in three dimensions, i.e., $x, y$, and $z$. 
Given an input matrix $V$, NMF produces a basis matrix $W \in \mathbb{R}^{P\times K}$, which means that the basis features of the vertices are indexed by $K$, and the coefficient $H \in \mathbb{R}^{K\times3}$ is indexed by the vertex positions. 
After obtaining several W, we selected K basis matrices from them that have the most semantic features.



\subsection{Latent Space Manipulation} \label{sec:latent}
\subsubsection{Projection Matrix Layer}
Our encoder takes a whole shape as input and compresses to a low-dimensional representation, i.e., a latent vector. This encoding reflects the whole shape structure. When we factorize the whole encoding, we can generate part encodings corresponding to the shape structure of the part. Thereby, we disentangle different semantic part encodings from the encoding of the whole shapes. We then perform part-level shape manipulation.

Dubrovina et al.~\cite{dubrovina2019composite} use projection matrices to transform a whole shape embedding into semantic part embeddings. They factorize the latent space into a semantic subspace with data-driven learned parameters. Similarly to ~\cite{dubrovina2019composite}, we use learnable projection matrices to transform the whole part encoding from the global latent space to the localized basis matrix space. We define part-specific projection matrices, where $K$ is the number of semantic parts. Passing through the matrices, the whole part encodings from the encoder are divided into semantic part encodings. 


For embedding parts, we implement projection matrices represented as $K$ fully connected layers without biases and with the latent dimension size of $Z\times Z$. The input of the projection layers is a whole face encoding produced by the encoder, and their outputs are $K$ part encodings. The $K$ part encodings can be split unpredictably and have arbitrary meanings. To make them more semantically meaningful, we apply pre-computed local weights. We explain this in the following paragraph.

\subsubsection{Applying Local Weights to Factorized Part Encodings}
Li et al.~\cite{li2020sketch}  proposed sketch, a combination of random noise and features of the original data, produced by transforming vectors from the noise space to a basis matrix space in NMF. 

\begin{figure}[ht]
\centering
\includegraphics[width=\linewidth]{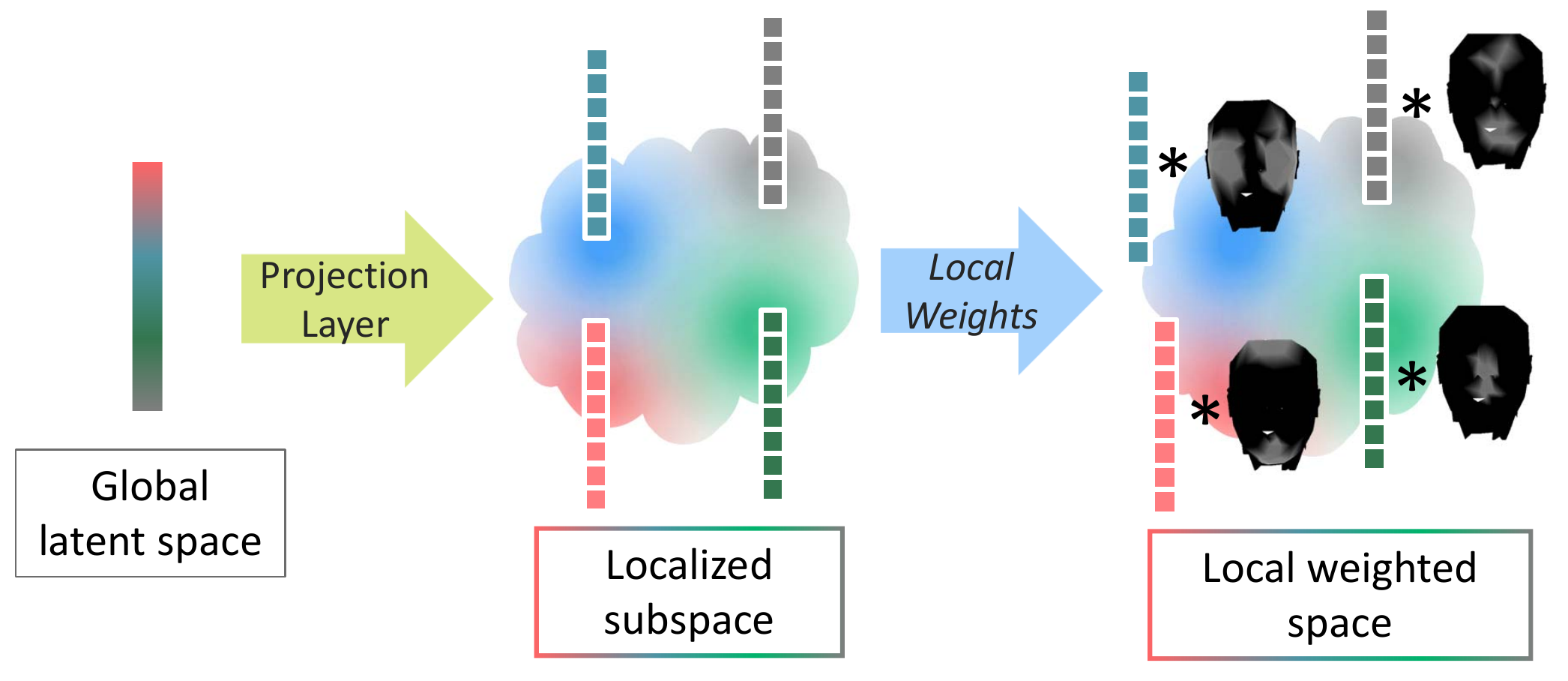}
\caption{
Illustration of the projection part}
\label{fig:projection}
\vspace{-10pt}
\end{figure}

Following~\cite{li2020sketch}, we apply the pre-computed local weights to the part encodings that are factorized by the projection matrices (Fig.~\ref{fig:full_arch}-(b)). Each pre-computed local weight is multiplied by each latent vector. Thanks to this operation, each factorized latent vector has a localized weight, and the encodings lie on a part-based subspace. We describe this process schematically in Fig.~\ref{fig:projection}.

Pre-computed local weights, which form the basis matrix derived from NMF $W\in\mathbb{R}^{P\times K}$, are applied to the factorized latent vectors $Z\in\mathbb{R}^{K\times Z}$. This process produces a locally weighted matrix $W_L\in\mathbb{R}^{K\times N\times Z}$ and then sum up as $W_L\in\mathbb{R}^{N\times Z}$ (Fig. ~\ref{fig:full_arch}). This matrix is provided as an input to the fully connected layer of the decoder.
\vspace{-10pt}

\section{Implementation}
\subsection{Datasets}
To obtain large facial mesh data, we used the AFLW2000-3D dataset~\cite{zhu2016face} having 2,000 3D faces and the corresponding landmarks of AFLW \cite{zhu2016face} face images. Each 3D face has 53,215 vertices. All faces are in full correspondence and generated by the Basel Face Model ~\cite{paysan20093d} without pose variations. In data pre-processing, we matched all facial mesh topology, i.e., those with the same vertex ordering. However, we did not clean some noisy area around the lips in the dataset, which is visible both in the dataset and in our synthetic results. The dataset was divided into a training set and a test set with 1,780 faces and 220 faces, respectively.

\subsection{Implementation Details}
Our proposed synthesis model has a similar architecture like CoMA~\cite{ranjan2018generating}, following their down-sampling and up-sampling method for coarse-to-fine convolution networks. The structure of the encoder and decoder is shown in Table~\ref{tb:encoder} and Table~\ref{tb:decoder}. Similar to COMA, our encoder contained four convolution layers, followed by a biased ReLU~\cite{glorot2011deep}. After passing the convolution layer, the input mesh was down-sampled approximately four times. The last fully-connected layer transformed the face mesh into a 64-dimensional latent vector. Once the first layer of the decoder transforms the latent vector, other processes mirror the encoder with an upsampling procedure, increasing the mesh data approximately four times. To optimize the networks, we exploit the $L_1$ and cycle loss~{\cite{dubrovina2019composite}}.

We trained our model for 300 epochs with a batch size of 32. The dimension of the latent vector was 64. The initial learning rate started at 0.0125 and decreased by 0.99 every epoch. We used stochastic gradient descent with a momentum of 0.9 to optimize and set Chebyshev filtering with K as 6. We used PyTorch~\cite{paszke2019pytorch} and PyTorch Geometric~\cite{fey2019fast} to implement our model and conducted all experiments with NVIDIA Titan RTX GPU 24GB. 

\begin{table}[h]
\centering
    \begin{tabular}{m{7em} m{4.5em} m{4.8em}}
        \hlineB{}
         Layer & Input size & Output size \\
         \hline\hline
         Convolution & $53215 \times 3$ & $53125 \times 16$ \\
         Down-Sampling & $53215 \times 16$ & $13304 \times 16$\\
         Convolution & $13304 \times 16$ & $13304 \times 16$\\
         Down-Sampling & $13304 \times 16$ & $3326 \times 16$\\
         Convolution & $3326 \times 16$ & $3326 \times 16$\\
         Down-Sampling & $3326 \times 16$ & $832 \times 16$\\
         Convolution & $832 \times 16$ & $832 \times 32$\\
         Down-Sampling & $832 \times 32$ & $208 \times 32$\\
         Fully Connected & $208 \times 32$ & 64\\
         \hlineB{}
    \end{tabular}
    \caption{Encoder Architecture}
    \label{tb:encoder}
    \vspace{-5pt}
\end{table}
\begin{table}[h]
\centering
    \begin{tabular}{m{7em} m{4.5em} m{4.8em}}
        \hlineB{}
         Layer & Input size & Output size \\
         \hline\hline
         Fully Connected & $208 \times 64$ &  $208 \times 32$\\
         Up-Sampling & $208 \times 32$ & $832 \times 32$\\
         Convolution & $832 \times 32$ & $832 \times 16$\\
         Up-Sampling & $832 \times 16$ & $3326 \times 16$\\
         Convolution & $3326 \times 16$ & $3326 \times 16$\\
         Up-Sampling & $3326 \times 16$ & $13304 \times 16$\\
         Convolution & $13304 \times 16$ & $13304 \times 16$\\
         Up-Sampling & $13304 \times 16$ & $53215 \times 16$\\
         Convolution & $53215 \times 16$ & $53125 \times 3$\\
         \hlineB{}
    \end{tabular}
    \caption{Decoder Architecture}
    \label{tb:decoder}
    \vspace{-10pt}
\end{table}


\section{Experimental Results}
The experimental results of our proposed model are described in this section. We present the practicality of our model with generation tasks and an ablation study. In all experiments, we set the number of face parts, $K$, as 4.

\subsection{Generation Results}
\subsubsection{Part manipulation}
\begin{figure} [ht]
\centering
\subfigure[Source]{\includegraphics[height=2.2cm]{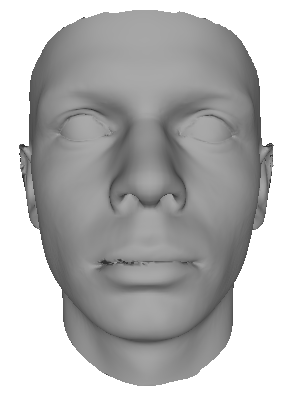}}
\hspace{1cm}
\subfigure[Target]{\includegraphics[height=2.2cm]{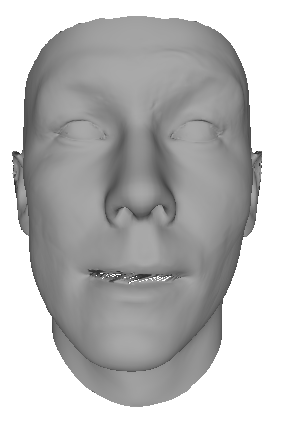}}
\caption{Source and target face}
\label{fig:snt}
\vspace{-10pt}
\end{figure}

\begin{figure*}[t]
\centering
\includegraphics [width=15cm]{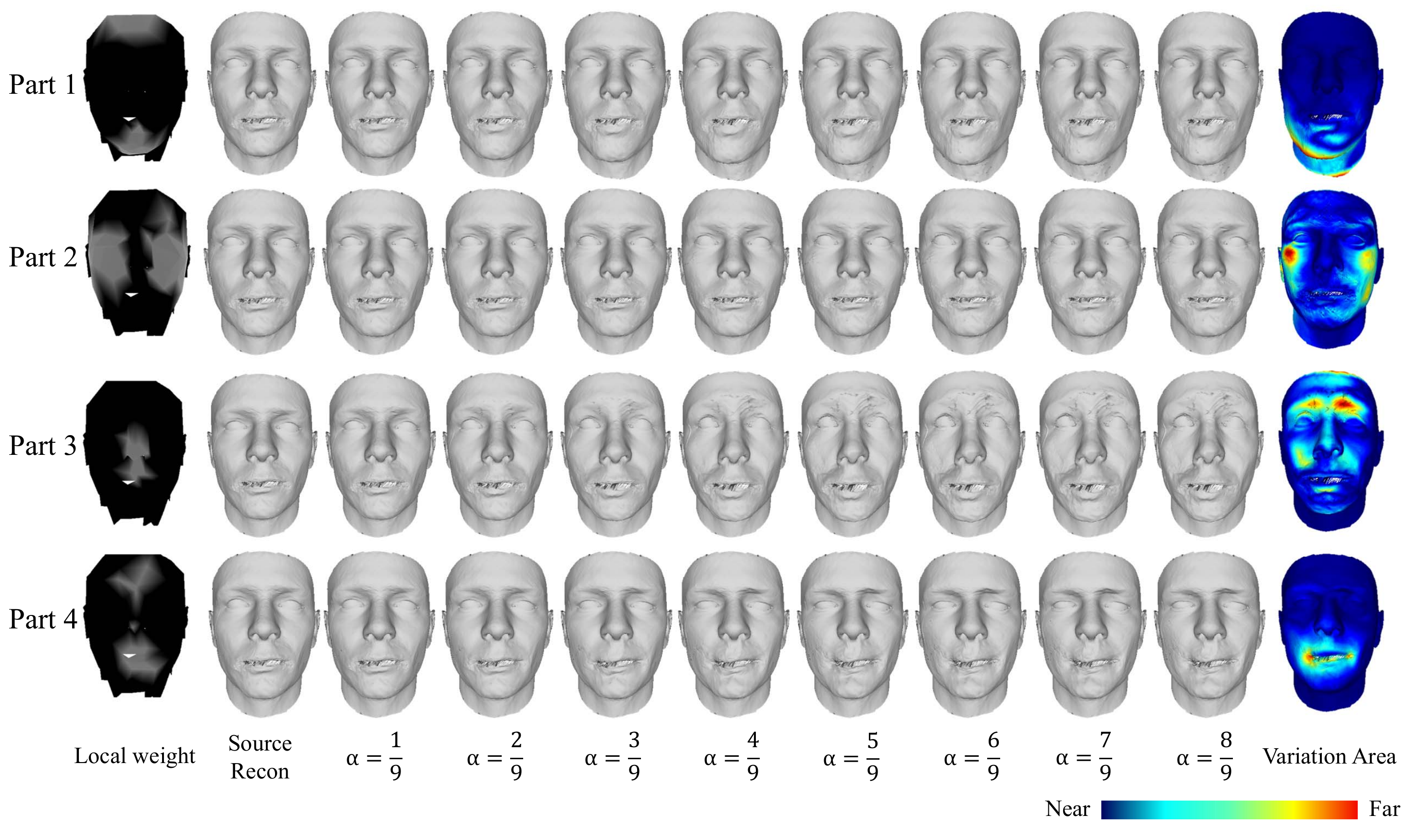}
\vspace{-10pt}
\caption{Results of part interpolation}
\label{fig:part1}
\end{figure*}

In this experiment, we tested the part manipulation results by applying interpolation between source and target as shown in Fig.~\ref{fig:snt}. We interpolated the source’s part encodings to the target’s corresponding part encodings obtained by factorized latent vectors described in Sec. \ref{sec:latent}. 
Fig.~\ref{fig:part1} shows that as the respective part of the face influence changes, the other parts of the face are not affected. Plus, we expected that each row’s changing part matches each local weight in the same row. As a result, we observed that each variation area corresponds to each local weight in Fig.~\ref{fig:part1}. Color gradients in the variation area included visualizing the Hausdorff distance between the first face $(\alpha=\frac{1}{9}\ )$ and the last face $(\alpha=\frac{8}{9}\ )$ in each row. Each of them displays a variation of each interpolation more clearly. The blue-colored gradient signifies that the vertices of the source and target are nearby, while the red-colored gradient means they are further away.

\subsubsection{Diversity Visualization}

To demonstrate the variety of data, we measured the diversity of generated data from our model.
Using the trained encoder, we encoded 220 random faces from our training set and test set, respectively. Since our proposed model allows part manipulation and modification, we synthesized 220 faces by combining five source faces and 11 target faces for four parts. Fig.~\ref{fig:syn} shows the synthesis samples with and without textures. 

\begin{figure} [bht]
\centering
\subfigure[PCA]{\includegraphics[width=3.5cm]{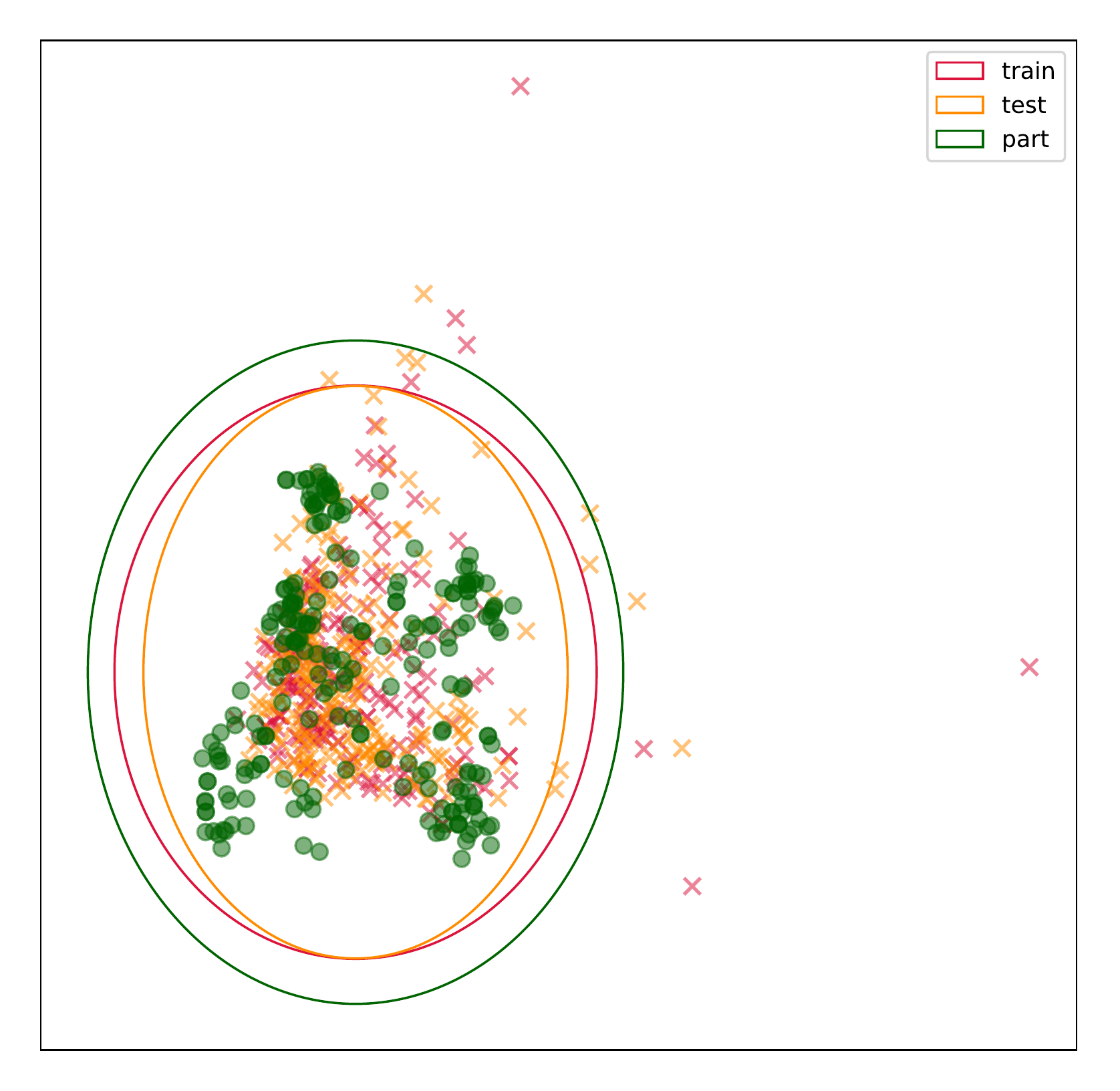}}
\subfigure[t-SNE]{\includegraphics[width=3.5cm]{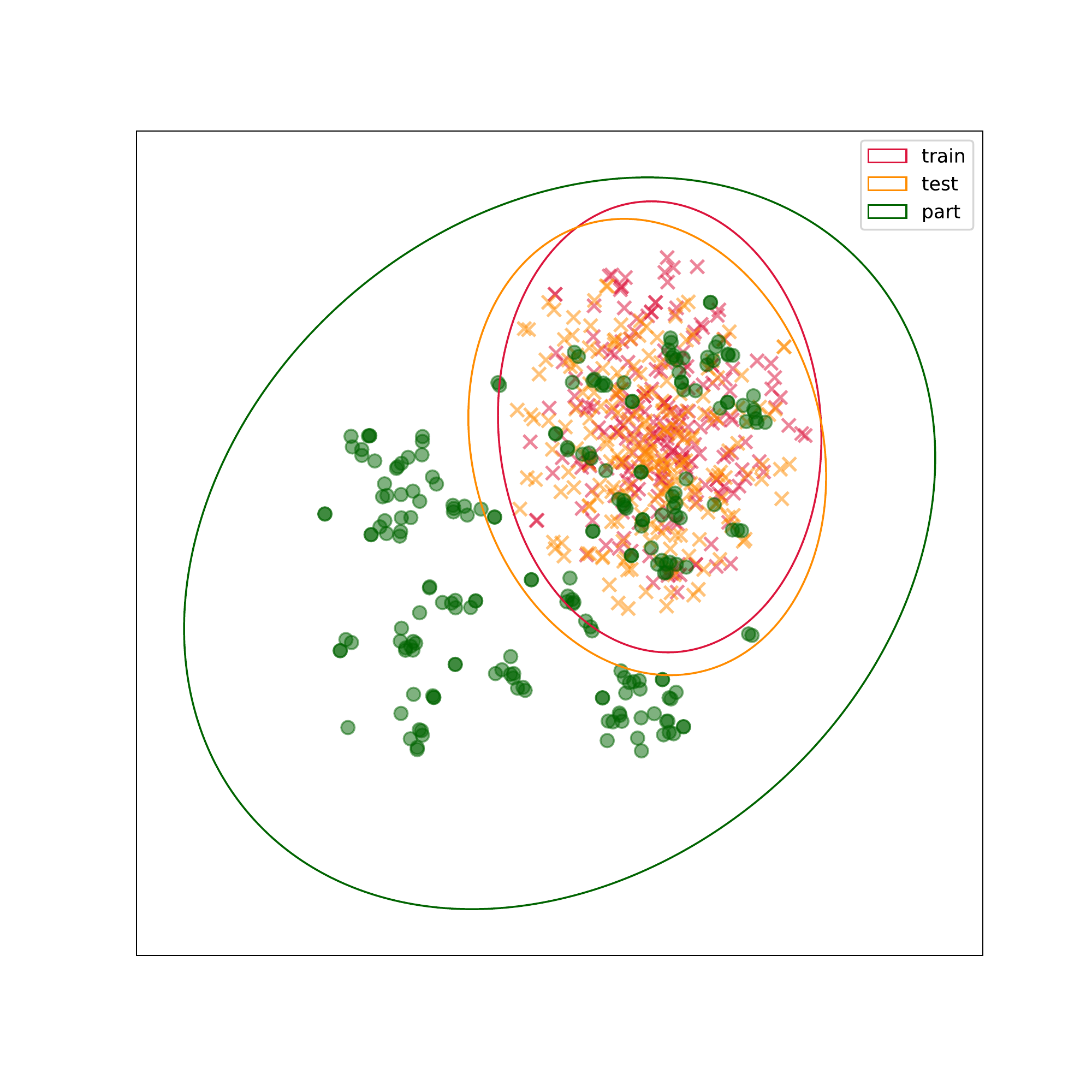}}
\caption{
Diversity Visualization}
\label{fig:vis}
\end{figure}

The result was visualized by projecting selected data onto a 2D plane using PCA and t-SNE~\cite{van2008visualizing}, shown in Fig.~\ref{fig:vis}. We displayed all encoded faces as markers and summarized them with ellipses. Here, there are three types of encoding: training set (red), test set (yellow), and part synthesis (green).


In Fig.~\ref{fig:vis} - (a), we can discern that our synthesis sample area (green ellipse) involves both areas of the training set and test set (red and yellow ellipses) in the 2D PCA plane. Fig.~\ref{fig:vis} - (b) presents this result more distinctively as the synthesis samples are also located in a wider region as well as the region of the training set and test set.
In our visualizations, even though the training data and test data overlap, our synthesis samples (green) cover wider areas in the encoding space. As a result, our proposed method shows a prominent performance to extend the model’s representation ability.

\subsection{Ablation Study}
To study the effect of each component of our approach, we experimented with an ablation study with variations of local weight and projection matrices.

\begin{figure*}
\centering
\includegraphics [width=13cm]{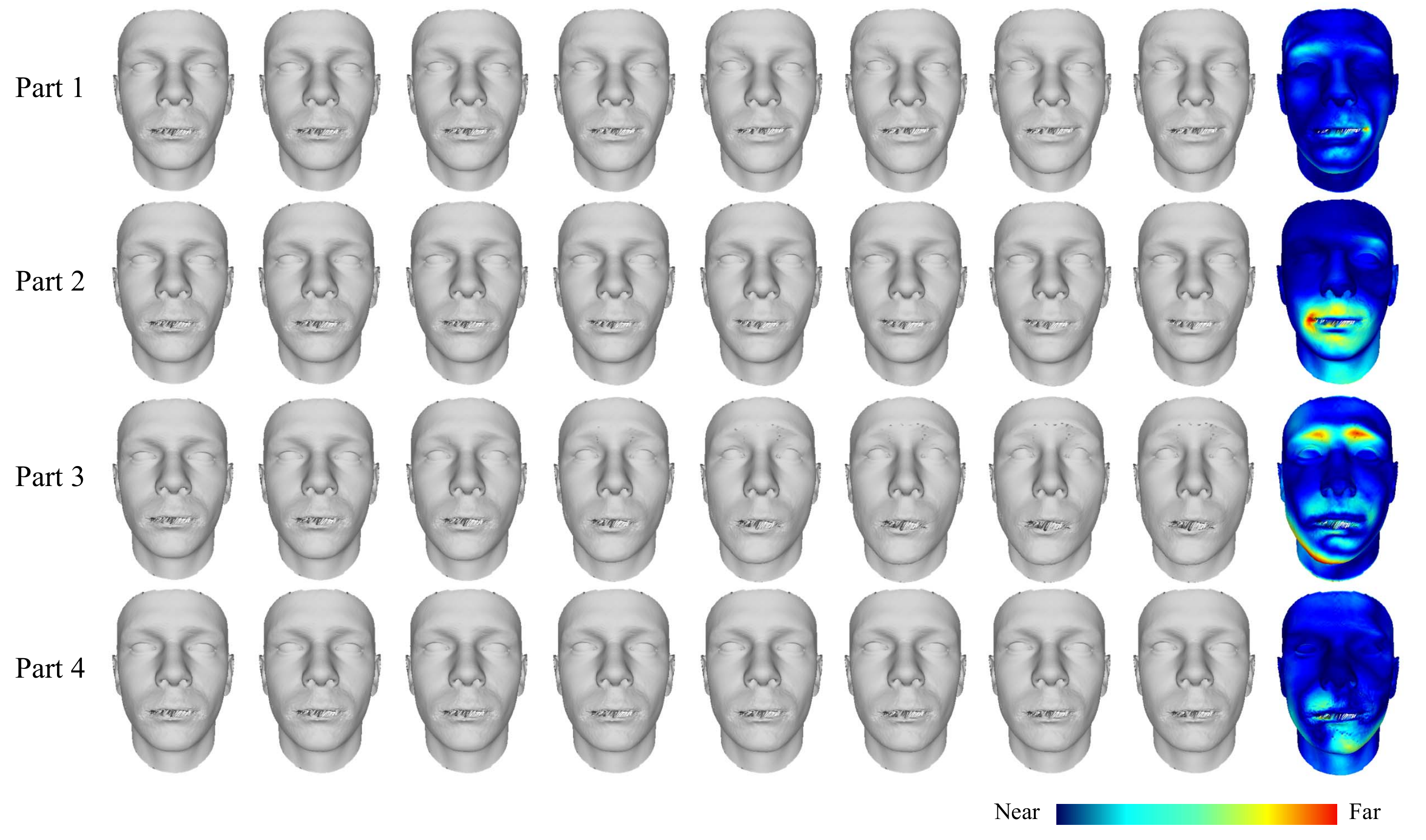}
\caption{
Results of part interpolation without applying local weights}
\label{fig:no_weight}
\end{figure*}

\begin{figure}
\centering
\includegraphics [width=\linewidth]{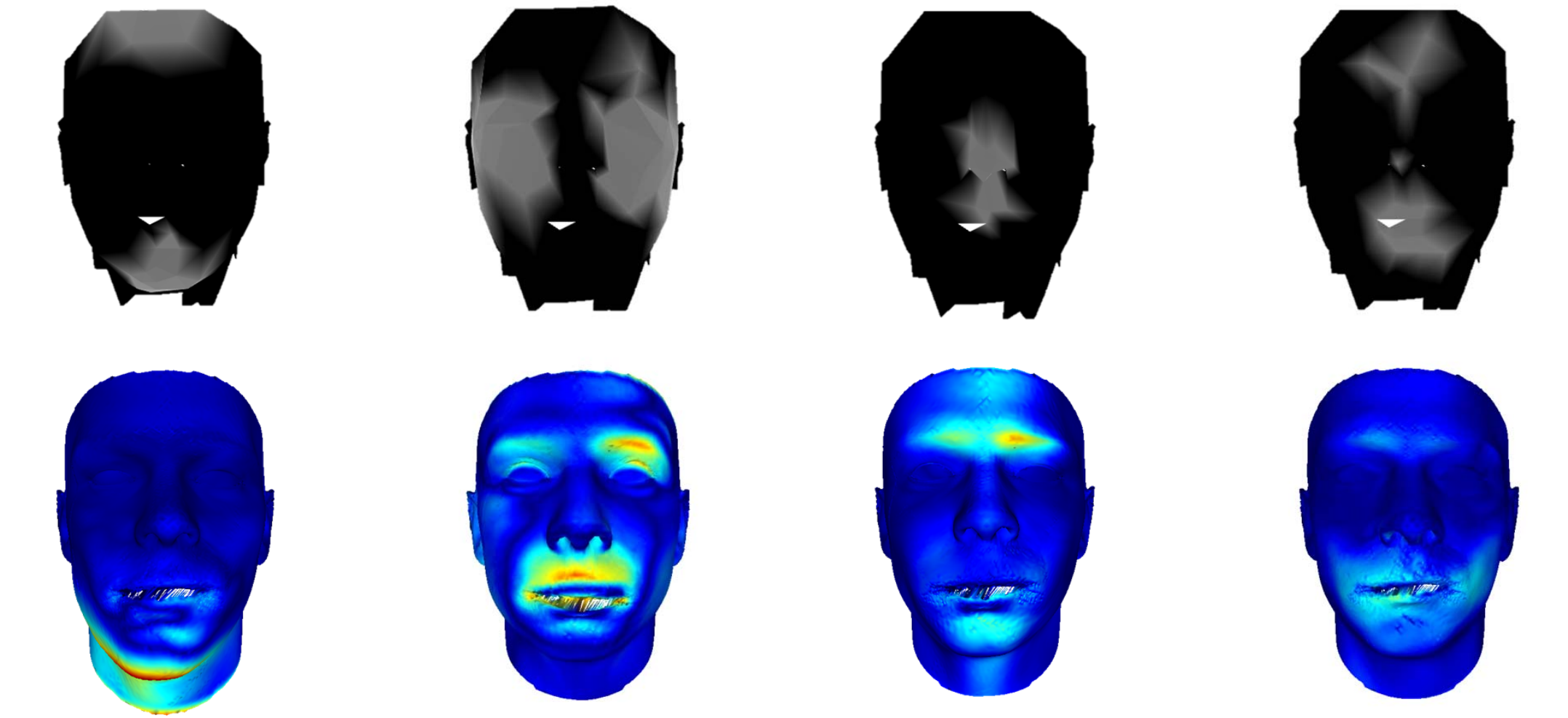}
\caption{
Part interpolation variation area results without factorization by projection matrices}
\label{fig:no_proj}
\vspace{-10pt}
\end{figure}

\subsubsection{Without local weights}
Local weights were obtained by NMF to make decomposed part spaces more semantically meaningful. To verify the effect of local weights, the model was trained without applying local weights to the projection part. The results are shown in Fig.~\ref{fig:no_weight} and are presented sequentially. The far-right face visualizes the Hausdorff distance between the first face and the last face in the sequence. Without applying local weight, the results do not display noticeable changes in some faces, and changing areas of the face also are intertwined with each other and look arbitrary.

\subsubsection{Without projection matrices}
In this experiment, we removed the factorization step to explore its impact. 
Compared to Fig.~\ref{fig:part1}, the changing parts of faces in Fig.~\ref{fig:no_proj} reflect less local weight except for the first face. Specifically, the second face’s cheek area impacts the mouth area, and other faces change different parts with local weight. Considering the results, we can infer that the projection matrices help exert local weights better. Accordingly, projection matrices not only factorize the latent space but also transform it into local weights space.

\subsection{Discussion}
Our proposed model performed notable part manipulation and synthesis using a holistic generative approach. However, there are a few points that need further discussion. 

Concerning the level of model components, the correlation between the changing area of faces and local weights should be better addressed. Most changing areas generally reflect corresponding local weights features, but some include another part or ignores them. One possible reason for this is that projection matrices would cover unassigned areas by transforming part encodings to local weight's space. The other is the natural quality of the dataset having correlations between facial features. Although we suggest two possible reasons, these need to be explored thoroughly. Next, the semantic meaning of local weights needs to be examined. Our local weights were computed algorithmically, not manually segmented or labeled. Therefore, it lacked semantic meaning and detailed segmentation of the human face, such as the separation of eyes and eyebrows. Finally, we multiply the part encodings in latent space and local weights in NMF. This approach seemed to work in our setting because the projection matrices transform part encodings to local weights’ space. We have shown experimentally that our process works, but a more rigorous mathematical proof is still needed. 

\section{Conclusion}
We proposed a locally weighted 3D generative face model using spectral convolution networks for a 3D mesh. Our model show improved expressiveness by manipulating the local parts of a face without explicit mesh segmentation. 

In future work, we would like to extend our model to apply other generative models i.e., VAE or GANs, to improve output’s quality. Generating face textures with geometry also would express the quality of outcomes better. Besides, it would be worthwhile to study part-based representation to improve the proposed local weights to develop the model’s synthesis ability.




\end{document}